\documentclass[dvips]{article}
\usepackage{icrctc07}

\title{Observation of the galaxy M87 with VERITAS}
\shorttitle{Observation of M87 with VERITAS}
\authors{Pierre Colin$^{1}$ for the VERITAS collaboration$^{2}$}
\shortauthors{P. Colin et al.}
\afiliations{$^1$ Department of Physics, Univ. of Utah, 115 S 1400
E, Salt Lake City, UT 84112, USA\\
$^2$ for full collaboration list see G. Maier, ''VERITAS: Status and
latest results''}
\email{colin@physics.utah.edu}

\abstract{The giant radio galaxy M87 is the only extragalactic non-blazar object which has been detected as a source of
very high energy $\gamma$-rays. It represents a unique opportunity to study the phenomena of $\gamma$-ray
emission from a nearby AGN. In this paper we report preliminary results from the observations of M87
taken with the imaging atmospheric Cherenkov telescope array VERITAS in February, March and April 2007.
An excess of photons above an energy threshold of 250~GeV is measured with a statistical significance
of more than five standard deviations.}

\begin{document}
\maketitle

\section{Introduction}
M87 is a nearby giant elliptical galaxy ($\sim$16~Mpc) which lies
near the center of the Virgo galaxy cluster. It is a powerful
radio source (Virgo A), classified as a Fanaroff-Riley class I (FR
I) radio galaxy~\cite{ref1}. Its core is an AGN powered by a
supermassive black hole of about $3.2\times 10^9~\mathrm{M}_\odot$
\cite{ref2}, emitting the first-detected plasma jet~\cite{ref3}
which extends over several kiloparsecs. According to the unified
scheme~\cite{ref4} of BL Lacertae objects (BL Lac) and FR I radio
galaxies, M87 should be of the same astrophysical nature as BL Lac but
with its jet not pointing along the line of sight. The critical
sight angle between BL Lac and FR I galaxies is not well known.
Some models propose an M87 jet misalignement around
30$^\circ$~\cite{ref5_bis} but superluminal motion observed by
HST~\cite{ref5} in the M87 jet suggest an orientation within
19$^\circ$ of the line of sight. The TeV emission from M87
establishes an important link between M87 and BL Lac objects,
which may have implications for the unified scheme, since all
other extragalactic sources detected in the TeV regime
are BL Lac objects~\cite{ref16}.\\

M87 has been observed over a broad range of energies from radio
waves to $\gamma$-rays. Its jet is resolved in radio, optical and
X-ray regimes and shows similar morphologies at all wavelengths.
This emission is understood as the synchrotron radiation from high
energy electrons in the jet. The same electron population could
also be responsible for the TeV $\gamma$-ray emission through the
inverse Compton process~\cite{ref6}. Models involving
protons~\cite{ref7} can also explain the TeV emission.

In fact, M87 is also considered as a possible source of ultra-high-energy ($\sim$10$^{20}$~eV) cosmic rays
\cite{ref8}. As tracers of high-energy hadrons, very-high-energy $\gamma$-rays could provide
important clues to help determine the origin of extragalactic cosmic rays.\\

The first detection of M87 in the TeV regime was reported by the
HEGRA collaboration~\cite{ref9} with a statistical significance of
4.1 standard deviations. This was confirmed by the HESS
collaboration~\cite{ref10} which also reported rapid variability
and an unexpectedly hard spectrum. The imaging atmospheric
Cherenkov technique provides insufficient angular resolution
($\sim3$~arcmin) to resolve the M87 emitting region (core, jet...)
but the day scale variability measured by HESS suggesting a very
small region, most likely close to the core. The observation of
superliminal motion of extremely compact radio structures within
the HST-1 region has however recently provided indications the TeV
emission could originate from that region, 0.82'' ($60$~pc
projected) from the core~\cite{ref15}.

\section{Observation with VERITAS}

VERITAS is an array of four imaging atmospheric Cherenkov telescopes with 12~m-diameter mirror and
3.5$^\circ$ field of view camera. Telescopes details can be found in these proceedings~\cite{ref11}.
The array is located in southern Arizona
where M87
reaches more than 70$^\circ$ in elevation.

The observations in the direction of M87 were carried out during 51 hours from February to April 2007 at elevations
from 55$^\circ$ to 71$^\circ$.
All data were taken in ''Wobble'' mode, tracking M87 with a 0.5$^\circ$ offset
(successively north, south, east, west).
VERITAS was still under construction and most of the observations were performed with a partial array
of three telescopes. In order to check the source variability, we processed all the data in the same way,
using only the first three telescopes even when the fourth telescope was in use (6\% of the data set).
We selected usable data on the basis of the weather conditions and the raw trigger rate stability.
About 90\% of the data pass our quality selection cuts (cf. Table~1).

\begin{table*}
\begin{center}
\begin{tabular}{|c||c|c||c|c|}
\hline
Date  &\multicolumn{2}{c|}{Observation time}&\multicolumn{2}{c|}{Results}\\
\cline{2-5}
 &raw &selected&excess rate&significance\\
\hline\hline
   12-25 Febr. & 17.0~h & 15.5~h & 0.14 $\pm$ 0.04 /min& 4.2 $\sigma$ \\
   12-20 March & 23.5~h & 20.4~h & 0.07 $\pm$ 0.03 /min& 2.6 $\sigma$  \\
   07-19 April & 10.7~h &  8.3~h & 0.10 $\pm$ 0.05 /min& 2.2 $\sigma$  \\
\hline\hline
   TOTAL & 51.2~h & 44.2~h & 0.10 $\pm$ 0.02 /min& 5.1 $\sigma$  \\
\hline
\end{tabular}
\caption{Data set toward M87 taken with VERITAS in spring 2007}\label{table-Data}
\end{center}
\end{table*}

\section{Data Analysis}
The data have been analyzed using independent analysis packages
(see~\cite{ref12} for details on the analysis). All of these
analysis yield consistent results. The presented results
are obtained with the GrISU package~\cite{ref17}. This
analysis chain differs from those described in~\cite{ref12} by fixed
thresholds in the image cleaning (same thresholds for all pixels)
and by the background-rejection cuts.
In this analysis, the background-rejection cuts were optimized for the
detection of a point-like source with a luminosity of 10\% of the
Crab Nebula. The signal from the source (''ON'') is the number of
events with a reconstructed direction less than 0.14$^\circ$ from
the source direction. The backround rate (''OFF'') is estimated
using seven ''mirror'' regions which have the same area and the
same distance from the camera center (0.5$^\circ$) as the ''ON''
region (reflected region method~\cite{ref12}). The statistical significance
is calculated using the Li and Ma formula~\cite{ref13} (with
$\alpha = 1/7$).

This analysis, applied to the Crab Nebula data taken with the 3-telescope-array
configuration at $\sim$78$^\circ$ elevation in February 2007,
results in an excess of 5.9~$\gamma$/min for a background rate of
0.75~$\rm evt/min$ (Li \& Ma excess significance of
27.5~$\rm \sigma/\sqrt{h}$). In this preliminary paper, we express the
measured $\gamma$-ray flux in Crab units. Simulations and data
show the excess rate does not vary significantly ($<10$\%)
between $78^\circ$ and the M87 observation elevation.

\section{Preliminary results}
VERITAS detected an excess of 263 events from the direction of M87,
corresponding to a statistical significance of 5.1 standard deviations
(secondary analysis yield also a significance above 5 standard deviations).
The energy threshold of our analysis at the M87 observation
elevation has been estimated with Monte Carlo simulations at 250~GeV.
The time averaged excess rate corresponds to a $\gamma$-ray flux of
$1.7\%$ of the Crab nebula flux.

\begin{figure}
\begin{center}
\includegraphics[width=\columnwidth]{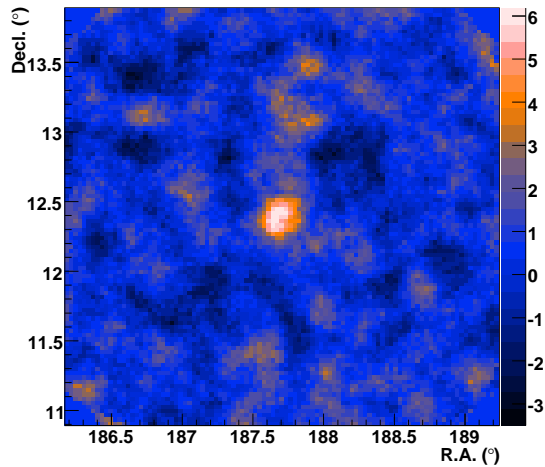}
\caption[short]{Excess significance map of the M87 surrounding region.}
\end{center}
\label{signi_map}
\end{figure}

\begin{figure}
\begin{center}
\includegraphics[width=\columnwidth]{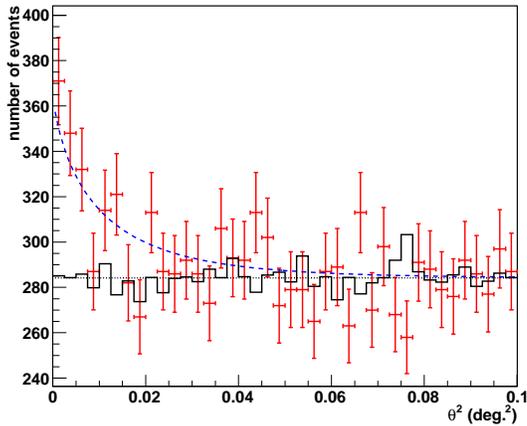}
\caption[short]{Crosses show the $\theta^2$ distribution from M87
and the solid line from off-source directions.
The dashed line is the $\theta^2$ distribution expected
for a point-like source with a backround level represented by the dotted line.}
\end{center}
\label{theta2}
\end{figure}

\begin{figure}
\begin{center}
\includegraphics[width=\columnwidth]{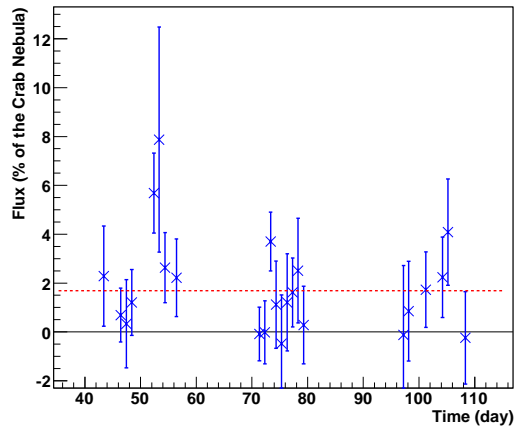}
\caption[short]{Nightly average (crosses) and global average (dashed line)
of the M87 $\gamma$-ray flux above 250~GeV
as a function of the observation day of year.}
\end{center}
\label{Wobble}
\end{figure}

\begin{figure}
\begin{center}
\includegraphics[width=\columnwidth]{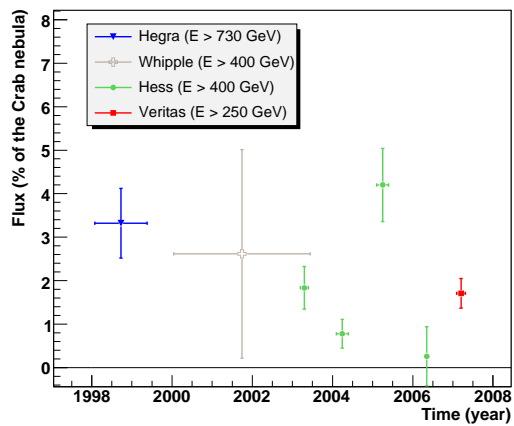}
\caption[short]{M87 flux reported by very high energy $\gamma$-ray telescopes
 as a function of observation year.}
\end{center}
\label{theta2}
\end{figure}

Figure~1 shows the map of the excess significance in the
surrounding region of M87. The excess position is: R.A. $12^h$
$30^m$ $47.5^s \pm 11^s$, Dec. $+12^\circ 23' 23'' \pm 1'48''$,
consistent with the position of M87 (R.A. $12^h$ $30^m$ $49.4^s$,
Dec. +12$^\circ$ 23' 28''). Figure~2 shows the distribution of the
square of the angle $\theta$ between M87 and the reconstructed
direction. The shape of the excess is compatible with a point-like
source (dashed line). The upper limit to the source extension has
a 2.3 arcmin radius.

Figure~3 shows the M87 light curve night-by-night during the three
months of observation. No significant variability is observed. The
constant flux fit $\chi^2$ per free parameters is: 21.4/22. The
maximal deviation of the rate from the average (February 21 and
22) reaches only 3.4 standard deviations. We cannot yet confirm
the variability on a 2-day scale reported by the HESS collaboration.

Figure~4 shows the M87 flux recorded during the last 10 years by
HEGRA~\cite{ref9}, Whipple 10~m~\cite{ref14}, HESS~\cite{ref10}
and VERITAS expressed in Crab units. The 2007 flux reported in
this paper is at an average level.

\section{Summary and Conclusion}
VERITAS confirms the $\gamma$-ray emission from M87 above 250~GeV.
The measured flux is below 2\% of the Crab Nebula flux, comparable to
earlier observations with HESS and HEGRA. This result demonstrates
the capability of VERITAS for detecting a faint source in 50h of
observations.

\section{Acknowledgements}
VERITAS is supported by grants from the U.S. Department of Energy, the
U.S. National Science Foundation and the Smithsonian Institution, by
NSERC in Canada, by PPARC in the U.K. and by Science Foundation Ireland.

\bibliography{icrc0756}

\begin{thebibliography}{10}

\bibitem{ref9}
F.~{Aharonian} et~al. (HEGRA~coll.).
\newblock {}.
\newblock {\em A\&A}, 403:L1, 2003.

\bibitem{ref10}
F.~{Aharonian} et~al. (HESS~coll.).
\newblock {}.
\newblock {\em Science}, 314:1424, 2006.

\bibitem{ref5_bis}
G.~V. {Bicknell} and M.~C. {Begelman}.
\newblock {}.
\newblock {\em ApJ}, 467:597, 1996.

\bibitem{ref8}
P.L. Biermann~et al.
\newblock {\em Physics and Astrophysics of UltraHigh-Energy Cosmic Rays}.
\newblock M. Lemoine \& G. Sigl (Berlin: Springer), 2001.

\bibitem{ref5}
J.~A. {Biretta}, W.~B. {Sparks}, and F.~{Macchetto}.
\newblock {}.
\newblock {\em ApJ}, 520:621, 1999.

\bibitem{ref15}
C.C. {Cheug}, D.E. {Harris}, and L.~{Stawarz}.

\bibitem{ref3}
H.~D. {Curtis}.
\newblock {}.
\newblock {\em Publ. Lick Obs.}, 13:9, 1918.

\bibitem{ref12}
M.~{Daniel}~et al.
\newblock {The VERITAS standard data analysis}.
\newblock In {\em Proc. 30th ICRC, Merida}, 2007.

\bibitem{ref1}
B.~L. {Fanaroff} and Riley~J. M.
\newblock {}.
\newblock {\em MNRAS}, 167:31P, 1974.

\bibitem{ref17}
{GrISU(tah):}.
\newblock {http://www.physics.utah.edu/gammaray/GrISU/}.

\bibitem{ref16}
H.~{Krawczynski}~et al.
\newblock {Blazar observations with VERITAS}.
\newblock In {\em Proc. 30th ICRC, Merida}, 2007.

\bibitem{ref14}
S.~{LeBohec}~et al.
\newblock {}.
\newblock {\em ApJ}, 610:156, 2004.

\bibitem{ref13}
T.P. {Li} and Y.Q. {Ma}.
\newblock {}.
\newblock {\em ApJ}, 272:317, 1983.

\bibitem{ref2}
F.~{Macchetto}~et al.
\newblock {}.
\newblock {\em ApJ}, 489:579, 1997.

\bibitem{ref11}
G.~{Maier}~et al.
\newblock {VERITAS: Status and latest results}.
\newblock In {\em Proc. 30th ICRC, Merida}, 2007.

\bibitem{ref7}
A.~{Reimer}, R.~J. {Protheroe}, and A.~C. {Donea}.
\newblock {}.
\newblock {\em A\&A}, 419:89, 2004.

\bibitem{ref6}
L.~{Stawarz}, L.~{Sikora}, and M.~{Ostrowski}.
\newblock {}.
\newblock {\em ApJ}, 597:186, 2003.

\bibitem{ref4}
C.~M. {Urry} and P.~{Padovani}.
\newblock {}.
\newblock {\em PASP}, 107:803, 1995.

\end{thebibliography}
\bibliographystyle{plain}

\end{document}